\documentclass[journal]{IEEEtran}

\ifCLASSINFOpdf
\else
   \usepackage[dvips]{graphicx}
\fi
\usepackage{url}
\usepackage{graphicx}
\usepackage{amsmath,etoolbox}
\usepackage{amssymb}
\usepackage[table]{xcolor}
\usepackage[caption=false]{subfig}
\usepackage{multirow}
\usepackage{cite}
\usepackage{upgreek}
\usepackage{float}
\usepackage{stfloats}

\DeclareMathOperator*{\argmax}{arg\,max}

\begin{document}

\title{Centroid Angle Estimation of Multiple Scatterers\\Using Monopulse Radar with Frequency Diversity}

\author{Minyoung Hwang, Hyuncheol Park, \IEEEmembership{Senior Member, IEEE}, and Joohwan Chun, \IEEEmembership{Senior Member, IEEE}
\thanks{The authors are with The School of Electrical Engineering, Korea Advanced Institute of Science and Technology, Daejeon 34141, Republic of Korea (e-mail: robco@kaist.ac.kr; hcpark@kaist.ac.kr; chun@kaist.ac.kr).}}

\markboth{}
{Shell \MakeLowercase{\textit{et al.}}: Bare Demo of IEEEtran.cls for IEEE Journals}
\maketitle

\begin{abstract}
The monopulse technique determines the angle of a target by comparing signals from two narrow beams, yielding a precise angular estimate with low complexity.
However, it struggles to resolve multiple closely spaced scatterers within the same resolution cell.
Existing methods for estimating multiple scatterer angles involve complex signal processing and system modifications.
We propose an effective method to estimate the angular centroid of scatterers using the mode of monopulse angle estimates.
A semi-analytic expression for the angle estimate distribution is derived, confirming that its mode aligns with the centroid.
To enhance estimation accuracy, we employ frequency diversity to reduce sample correlation.
Numerical results validate the advantages of the proposed method, demonstrating superior performance over conventional techniques with low complexity.
\end{abstract}


\IEEEpeerreviewmaketitle

\section{Introduction}
Radar usually employs a high-gain transmit beam (or antenna) to increase the signal-to-noise ratio (SNR) by focusing its power into a narrow beam footprint.
This, in turn, necessitates scanning the beam—either electronically or mechanically—to cover a wide surveillance region.
To detect a target and estimate its angle within a narrow transmit beam footprint, it is not necessary to apply computationally intensive optimal array signal processing techniques\cite{2002van} to the received echo signal in the element space of multiple antennas.
The monopulse technique estimates the angle of arrival of a target within the transmit beam footprint by simply comparing the amplitudes or phases of the signals received through a pair of narrow receive beams (or antennas)\cite{2011sherman}.
Notably, the Cramér–Rao bound of the monopulse angle estimate approaches that of the element-space estimate when the target is near the boresight of the receive beams\cite{2001nielsen}.
Consequently, the monopulse angle estimator is widely adopted in scanning military radars due to its computational simplicity and accuracy.

However, conventional monopulse processing assumes the presence of a single target and struggles to resolve multiple targets within the same range-Doppler resolution cell.
This situation frequently arises in a shipborne radar tracking a low-flying target over the sea surface\cite{2019oh,1991zoltowski,2014park}, in a radar seeker approaching an extended stationary target, and—more pertinently to this study—in a close-in weapon system (CIWS)\cite{1993serakos}.
In CIWS applications, monopulse radar is used to estimate the centroid angle of outgoing friendly bullets, enabling the gun barrel to be adjusted to align with an incoming target\cite{1993serakos}.

The problem of estimating the individual angles of multiple coherent scatterers using an antenna array—either in the element space or the beam space—has been extensively studied, for example in \cite{2002van,2019oh,1985shan,1991zoltowski,2015lee}.
The iterative method of nulling the multipath signal of a low-angle target proposed in \cite{2014park} addresses two coherent scatterers, but can be readily extended to handle more than two and estimate their individual angles.
If we have estimates of the individual angles, the centroid angle can be found simply by averaging them.
Furthermore, the direct estimation of the centroid angle of multiple scatterers—as well as their angular spread—has also been investigated, both in the element space of an antenna array\cite{1996trump} and in the beam space, assuming cosine and sine patterns for the sum and delta beams, respectively\cite{2012monakov}.
However, all the aforementioned algorithms for estimating the centroid angle are computationally intensive and require extensive modifications to the existing conventional monopulse radar systems.
Some studies have investigated the decision problem of determining whether one or more scatterers are present within the resolution cell, as well as the angle estimation of the primary scatterer, using conventional monopulse radar\cite{2011sherman,1986bogler}.
However, these efforts have largely focused on mitigating the effects of secondary reflections rather than estimating the centroid angle itself.

In this work, building upon our preliminary study in \cite{2023chun} that introduced the concept of mode-based estimation, we propose a comprehensive framework for monopulse centroid estimation.
We first demonstrate that the mode of conventional monopulse angle estimates, when applied to multiple scatterers, closely aligns with the angular
centroid of those scatterers.
To the best of our knowledge, this result has not been reported in the open literature.
Furthermore, to reduce the variance of the mode-based estimate, we introduce frequency diversity, which lowers the correlation between measurements taken at different frequencies.
The proposed method allows a conventional monopulse radar system to estimate the centroid angle of multiple scatterers—or an extended target—without requiring any architectural changes or software modifications.
Moreover, the proposed method outperforms conventional centroid estimation techniques\cite{1996trump,2012monakov} in sparse scattering environments while maintaining computational efficiency.

\section{Centroid Angle Estimation with Monopulse Radar}\label{sec:system}
\subsection{Monopulse Operation}
We consider an array with $N_a$ antennas spaced at a distance of $\lambda/2$, operating at a frequency of $f_0$.
The separation between the phase centers of two beams is $d=8\lambda$.
Noiseless measurements on a radar signal reflected from a point target located at an angle $\theta$ are denoted as $z_0=we^{-j\phi}$ and $z_1=we^{j(u-\phi)}$, where $u=kd\sin{\theta}$, $k=2\pi/\lambda$ and $\lambda$ is the wavelength of the signal.
The magnitude and phase of the reflected signal are $w$ and $\phi$, respectively.
Next, sum and delta signals of the measurements are defined as $\Sigma=z_0+z_1$ and $\Delta=z_0-z_1$, respectively.
The monopulse ratio of sum and delta signals is given as follows\cite{2011sherman}.
\begin{equation}\label{eq:monopulse_ratio}
    A=\frac{\Delta}{\Sigma}=\frac{z_0-z_1}{z_0+z_1}=\frac{1-e^{ju}}{1+e^{ju}}=-j\tan{\frac{u}{2}}.
\end{equation}

The target angle can be inferred from the monopulse ratio in \eqref{eq:monopulse_ratio} using the following equation.
\begin{equation}\label{eq:estimates}
    u=2\tan^{-1}{\left(\Im\left\{-A\right\}\right)},\,\theta=\sin^{-1}{\left(\frac{u}{kd}\right)},
\end{equation}
where $\Im\{c\}$ denotes the imaginary part of a complex number $c$.
It is known that the angle obtained from \eqref{eq:estimates} is approximately the maximum likelihood estimation for noisy measurements\cite{2006nickel}.

\subsection{Monopulse Estimation of Multiple Scatterers with Frequency Diversity}\label{sec:estimation}
Suppose a monopulse radar transmits $N$ pulses in a single burst with a pulse repetition interval of $T$, towards an ensemble of $M$ unresolved scatterers.
The distance to the $m$-th scatterer at the moment of the $n$-th pulse is
\begin{equation}
    r_{n,m}=r_{1,m}+\dot{r}_m(n-1)T,\quad n=1,\cdots,N,
\end{equation}
where $r_{1,m}$ is the distance to the $m$-th scatterer at the initial pulse and $\dot{r}_m$ is the radial velocity of the scatterer.
The phase of the $n$-th pulse reflected on the $m$-th scatterer is $\phi_{n,m}=2\pi f_0 \tau_{n,m}$ where $\tau_{n,m}=2r_{n,m}/c$ is the round-trip delay.

The noisy measurements of two beams for the $n$-th pulse reflected on $M$ scatterers are given as
\begin{equation}\label{eq:measurements}
    \mathbf{z}[n]=\begin{bmatrix}
        1        & 1        & \cdots & 1\\
        e^{ju_1} & e^{ju_2} & \cdots & e^{ju_M}
    \end{bmatrix}\mathbf{s}[n]+\boldsymbol{\omega}[n]\in\mathbb{C}^{2\times1},
\end{equation}
where $\mathbf{s}[n]=\left[w_1e^{j\phi_{n,1}},\cdots,w_Me^{j\phi_{n,M}}\right]^T$, $w_m$ is the magnitude of the signal reflected from the $m$-th scatterer, and $\boldsymbol{\omega}[n]$ is zero-mean complex Gaussian additive noise with covariance $\sigma_\omega^2\mathbf{I}_2$.
Monopulse radar is unable to resolve the angles of individual scatterers because the signals reflected from each scatterer are received as a superposition.
In practice, monopulse radar can estimate the angular centroid of the scatterer ensemble.
We found that the mode of angles estimates derived from measurements in \eqref{eq:measurements} yields the angular centroid of the scatterers.

The estimate of the centroid angle is accurate when the phases of scatterers are uncorrelated between samples.
However, the phases of scatterers can have a strong correlation when the scatterers tend to have identical speeds within an ensemble.
For example, counter bullets of CIWS tend to have identical speeds, which can lead to difficulty in accurate centroid estimation.
Therefore, we propose the use of frequency diversity to enhance the accuracy of centroid estimation of multiple scatterers.
One straightforward method to implement frequency diversity is by varying the frequency of each pulse within a burst.
For simplicity, we consider stepped frequency diversity as an example, where the phase of the $n$-th pulse reflected from the $m$-th scatterer is given by
\begin{equation}\label{eq:phase_freq_div}
    \phi_{n,m}=2\pi f_n \tau_{n,m},\quad f_n=f_0+(n-1)\Delta f,
\end{equation}
where $\Delta f$ is the frequency increment between successive pulses.
Even if the radial velocities of scatterers are identical, i.e., $\tau_{n,m}$ is static over $n$, the phase will have diversity due to varying frequency.

\section{Distribution of the Estimation Statistic}\label{sec:statistic}
We derive a semi-analytical expression for the distribution of angle estimates for multiple scatterers.
Next, we demonstrate that the mode of the derived distribution aligns with the weighted mean of the angle estimates, which is the angular centroid of the scatterers.
For convenience, the distribution is first derived with respect to $u$ rather than the angle $\theta$.
As shown in \eqref{eq:estimates}, $\theta$ can be represented as a monotonic one-to-one function of the variable $u$, thus, the transformation of the distribution of $u$ into that of $\theta$ is straightforward.

The measurements of two monopulse beams, when the number of scatterers is $M$, are denoted as
\begin{equation}
    z_0=\sum_{m=1}^M{w_me^{-j\phi_{m}}},\quad z_1=\sum_{m=1}^M{w_me^{j(u_m - \phi_{m})}},
\end{equation}
where $w_m$ is a deterministic nuisance parameter and $\phi_m$ is a random nuisance parameter that depends on the distance between the radar and the scatterer.
We assume that the phase of each scatterer is independent and uniformly random, i.e., $\phi_m\sim U(-\pi,\pi)$, throughout the derivation.
This can be achieved using frequency diversity in practice.
Next, an angle estimate function is defined as the angular difference of two measurements taken on $M$ closely spaced scatterers.
\begin{equation}\label{eq:angle_diff}
    u=f(\boldsymbol{\Phi})=\arg{(z_1)}-\arg{(z_0)},\quad\boldsymbol{\Phi}=[\phi_1,\phi_2,...,\phi_M]^T.
\end{equation}
The domain of the function \eqref{eq:angle_diff} is
\begin{equation}
    D=\{\boldsymbol{\Phi}|-\pi\leq\phi_m\leq\pi,\,sp(\boldsymbol{\Phi})=r\},
\end{equation}
where the spread of $\boldsymbol{\Phi}$ is defined as follows.
\begin{equation}
    sp(\boldsymbol{\Phi})=\max_{1\leq n,m \leq M}{|\phi_n - \phi_m|}.
\end{equation}
The probability density of a random variable $u=f(\boldsymbol{\Phi})$ for $0\leq r\leq R$ is derived as
\begin{equation}\label{eq:pdf}
    p_{u}(u|sp(\boldsymbol{\Phi})\leq R) = \int_{0}^{R} p_{u}(u|sp(\boldsymbol{\Phi}) = r)p(sp(\boldsymbol{\Phi}) = r)dr.
\end{equation}

We first consider the probability density for the $M=3$ case.
The three-dimensional domain of the function \eqref{eq:angle_diff} can be partitioned into six disjoint one-dimensional (1-D) domains, which are expressed as
\begin{equation}
    D_i = \{\phi_{\sigma_i(1)} = 0, 0\leq\phi_{\sigma_i(2)}\leq r,\phi_{\sigma_i(3)}=r\},\,i=1,...,6,
\end{equation}
where $\sigma_i$ represents the $i$-th permutation of an array of orders $(1,2,3)$, e.g., $\sigma_1=(1,2,3)$, $\sigma_2=(1,3,2)$, $\sigma_3=(2,1,3)$, $\sigma_4=(2,3,1)$, $\sigma_5=(3,1,2)$, and $\sigma_6=(3,2,1)$.
The partitioned function with the domain $D_i$ is defined as
\begin{align}
    f_i(\phi_{\sigma_i(2)}=\phi)=&\arg\left( w_1^ie^{ju_1^i}+w_2^ie^{j(u_2^i-\phi)}+w_3^ie^{j(u_3^i-r)} \right)\nonumber\\
    -&\arg\left( w_1^i+w_2^ie^{-j\phi}+w_3^ie^{-jr} \right),
\end{align}
where $w_m^i=w_{\sigma_i(m)}$ and $u_m^i=u_{\sigma_i(m)}$ for $m\in\{1,2,3\}$.

Using the transformation theorem on the function of random variables\cite{2009gut}, the density function of $u$ is derived as follows.
\begin{equation}\label{eq:pdf_u}
    p_u(u|sp(\boldsymbol{\Phi})=r)=\sum_{i=1}^{6}{p_u^{(i)}(u|sp(\boldsymbol{\Phi})=r)},
\end{equation}
\begin{equation}\label{eq:pdf_i}
    {p_u^{(i)}(u|sp(\boldsymbol{\Phi})=r)}={\sum_{j=1}^{K}\left|\frac{df_{i,j}^{-1}(u)}{du}\right|p_{\phi}(f_{i,j}^{-1}(u)|sp(\boldsymbol{\Phi})=r)},
\end{equation}
where $f_{i,j}$ denotes the $j$-th function in the decomposition of $f_i$ into $K$ monotonic injective functions.
The derivative of the inverse of $f_{i,j}$ with respect to $u$ can be calculated numerically.
As an alternative approach, we approximate $f_{i,j}$ as a linear function passing through $f_{i,j}(r_{i,j-1})$ and $f_{i,j}(r_{i,j})$ where $r_{i,j}$ is the $j$-th turning point of $f_i$.
Note that $r_{i,0}=0$ and $r_{i,K}=r$.
Consequently, the probability density in \eqref{eq:pdf_i} simplifies to
\begin{align}\label{eq:f_lin}
    &p_u^{(i)}(u|sp(\boldsymbol{\Phi})=r) \\
    &= \begin{cases}
    1/|f_{i,j}(r_{i,j})-f_{i,j}(r_{i,j-1})|, & \text{if $u_{\min}^{(i,j)} \leq u \leq u_{\max}^{(i,j)}$}, \\
    0, & \text{otherwise}, \nonumber
  \end{cases}
\end{align}
where $u_{\min}^{(i,j)}=\min(f_{i,j}(r_{i,j-1}),f_{i,j}(r_{i,j}))$, and $u_{\max}^{(i,j)}=\max(f_{i,j}(r_{i,j-1}),f_{i,j}(r_{i,j}))$.

Let $\Phi_{(1)}$, $\Phi_{(2)}$, $\Phi_{(3)}$ denote the order statistics of uniformly random phase samples.
The spread $r=\Phi_{(3)}-\Phi_{(1)}$ is the difference between the smallest and largest phase measurements.
Thus, the distribution of spread is computed as \cite{2002casella}
\begin{equation}\label{eq:pdf_r}
    p(sp(\boldsymbol{\Phi}) = r)=\frac{3}{4\pi^3}(2\pi r-r^2),
\end{equation}
for $0\leq r\leq2\pi$.
Finally, substitution of \eqref{eq:pdf_u}, \eqref{eq:f_lin}, and \eqref{eq:pdf_r} into \eqref{eq:pdf} yields the distribution of angle estimate.

\begin{figure}
\centerline{\includegraphics[width=.5\columnwidth]{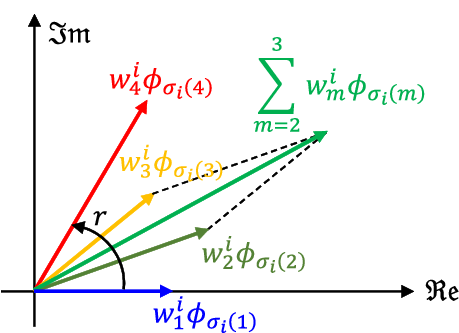}}
\caption{Conversion of $M=4$ ensemble into $M=3$ ensemble.}
\label{fig:conversion}
\end{figure}

Now, we generalize the distribution derivation to the case where $M$ is greater than 3.
In order to decompose the domain of the angle estimate function into 1-D domains, we combine the responses to convert the ensemble into one composed of 3 scatterers effectively.
For each $i$-th permutation among $M!$ permutations, we fix $r=\phi_{\sigma_i(M)}-\phi_{\sigma_i(1)}$ and consider the sum of $w_m^i e^{-j\phi_{\sigma_i(m)}}$ for $m=2,...,M-1$ as a single response.
The $i$-th domain of the function \eqref{eq:angle_diff} is
\begin{equation}
    D_i=\{\phi_{\sigma_i(1)} = 0,\,\phi_{\sigma_i(M)}=r,\,0\leq[\boldsymbol{\Phi}_{\sigma_i}]_m\leq r,\,\forall m\},
\end{equation}
where $\boldsymbol{\Phi}_{\sigma_i}=[\phi_{\sigma_i(2)},...,\phi_{\sigma_i(M-1)}]^T$ and $[\mathbf{\cdot}]_m$ denotes the $m$-th element of the vector.
The angle estimate function with the domain $D_i$ is defined as follows.
\begin{align}\label{eq:f_part_general}
    &f_i(\boldsymbol{\Phi}_{\sigma_i})=\nonumber\\
    &\arg{\left(w^i_1e^{ju^i_1}+\sum_{m=2}^{M-1}{w^i_me^{j(u_m^i - \phi_{\sigma_i(m)})}}+w^i_Me^{j(u^i_M-r)}\right)}\nonumber\\
    &-\arg{\left(w^i_1+\sum_{m=2}^{M-1}{w^i_me^{-j\phi_{\sigma_i(m)}}}+w^i_Me^{-jr}\right)}.
\end{align}
Next, the probability density of $u$ is calculated by numerically integrating $\boldsymbol{\Phi}_{\sigma_i}$ over all possible phase combinations.
\begin{equation}\label{eq:pdf_u_general}
    p_u(u|sp(\boldsymbol{\Phi})=r)=\sum_{i=1}^{M!}\int_{\varphi}p_u^{(i)}(u|sp(\boldsymbol{\Phi})=r,\boldsymbol{\Phi}_{\sigma_i}=\boldsymbol{\upphi})d\boldsymbol{\upphi},
\end{equation}
where $\varphi=[0,r]^{M-2}$ denotes the range of phase elements, and $\boldsymbol{\upphi}\in\varphi$ denotes the $M-2$ phase elements in between $\phi_{\sigma_i(1)}$ and $\phi_{\sigma_i(M)}$.
The rest of the process to obtain the distribution of estimates becomes equivalent to the case with $M=3$.
Fig. \ref{fig:conversion} illustrates the example of converting $M=4$ to $M=3$.

\section{Numerical Results}\label{sec:simulation}
\begin{figure}
\centering
\subfloat[$M=3$]{\includegraphics[width=.5\columnwidth]{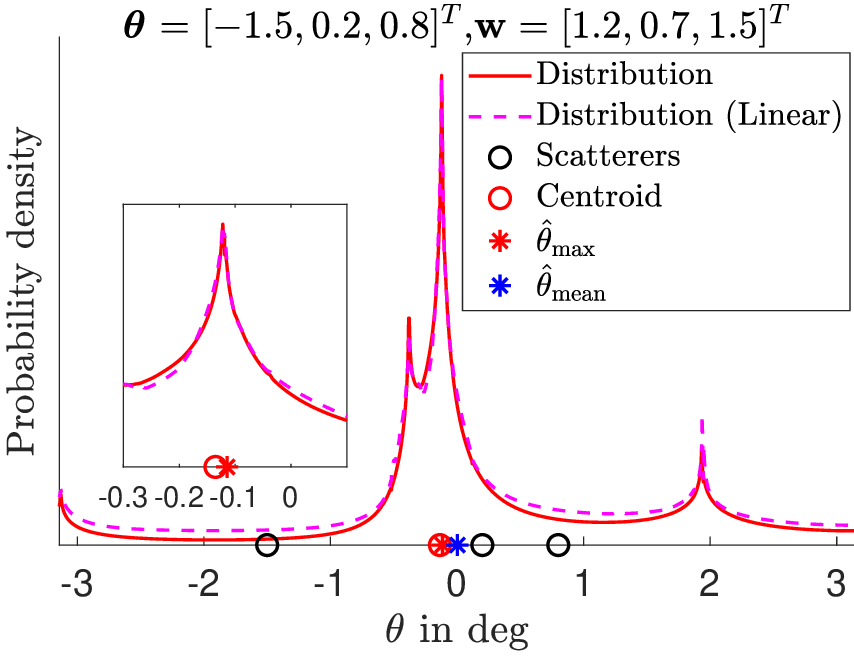}%
\label{fig:dist_Nt3}}
\subfloat[$M=4$]{\includegraphics[width=.5\columnwidth]{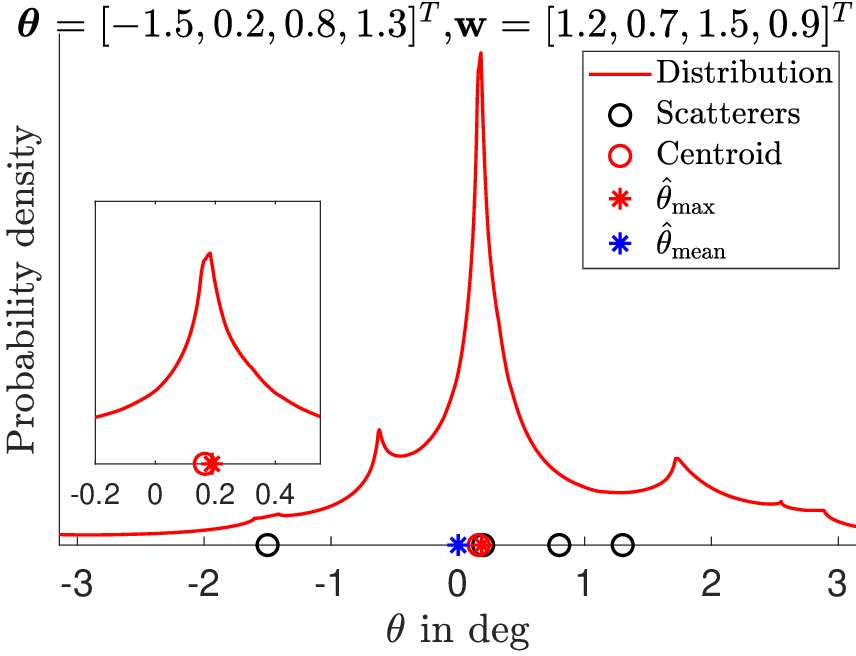}%
\label{fig:dist_Nt4}}
\caption{Distribution of the estimation statistic.}
\label{fig:dist}
\end{figure}

We present numerical results to verify that the mode of the distribution \eqref{eq:pdf} corresponds to the weighted mean of the angles of the scatterers as follows.
\begin{equation}\label{eq:u_max}
    \hat{\theta}_{\max}=\argmax_\theta{p_\theta(\theta|sp(\boldsymbol{\Phi})\leq R)}=\frac{\sum_{m=1}^{M}w_m \theta_m}{\sum_{m=1}^{M}w_m},
\end{equation}
where the probability density of $\theta$ can be obtained by transforming \eqref{eq:pdf} as in \eqref{eq:estimates} without loss of generality.
Fig. \ref{fig:dist} shows the distributions of estimated angles for targets composed of $M=3$ and $M=4$ scatterers.
The red line in Fig. \ref{fig:dist_Nt3} shows the computed distribution of the angle estimate in \eqref{eq:pdf} when $M=3$.
The dashed magenta line denotes the distribution computed using linear approximation in \eqref{eq:f_lin}.
Note that the mode of the distribution appears at the weighted mean of $\theta$, implying that $\hat{\theta}_{\max}\triangleq\mathrm{mode}(\theta)$ is a reliable estimator of the centroid of multiple scatterers.
This observation remains valid when $M=4$, where the mode continues to align with the centroid, as illustrated in Fig. \ref{fig:dist_Nt4}.
Since the distribution is skewed, the mode of the distribution does not generally coincide with the mean of the estimates, i.e., $\hat{\theta}_{\max}\neq\hat{\theta}_{\mathrm{mean}}\triangleq\mathrm{mean}(\theta)$.

\begin{table}
    \centering
    \begin{tabular}{|c|c|c|c|c|} 
        \hline
        \rowcolor{lightgray!20}
        $f_0$ [GHz] & $d$ [$\lambda$] & $r$ [m]          & $\theta$ [deg] & $M$               \\
        \hline
        10          & 8               & $U(-10,10)$      & $U(-1,1)$  & 4                 \\
        \hline
        \hline
        \rowcolor{lightgray!20}
        Fig         & $N$             & $\Delta f$ [MHz] & SNR [dB]       & $\dot{r}$ [m/sec] \\
        \hline
        \multirow{2}{*}{\ref{fig:rmse_snr}} & \multirow{2}{*}{32} & \multirow{2}{*}{0 or 10} & \multirow{2}{*}{$-20$ $\sim$ 30} & 1100 or \\
        &&&& $U(1090,1110)$
        \\
        \hline
        \multirow{2}{*}{\ref{fig:rmse_n}} & \multirow{2}{*}{1 $\sim$ 130} & \multirow{2}{*}{0 or 150$/N$} & \multirow{2}{*}{20} & 1100 or \\
        &&&& $U(1090,1110)$
        \\
        \hline
    \end{tabular}
    \caption{Parameters used for simulation results in Fig. \ref{fig:rmse}}
    \label{tab:params}
\end{table}

\begin{figure*}[!bp]
\centering
\subfloat[Comparison by SNR settings]{\includegraphics[width=.65\columnwidth]{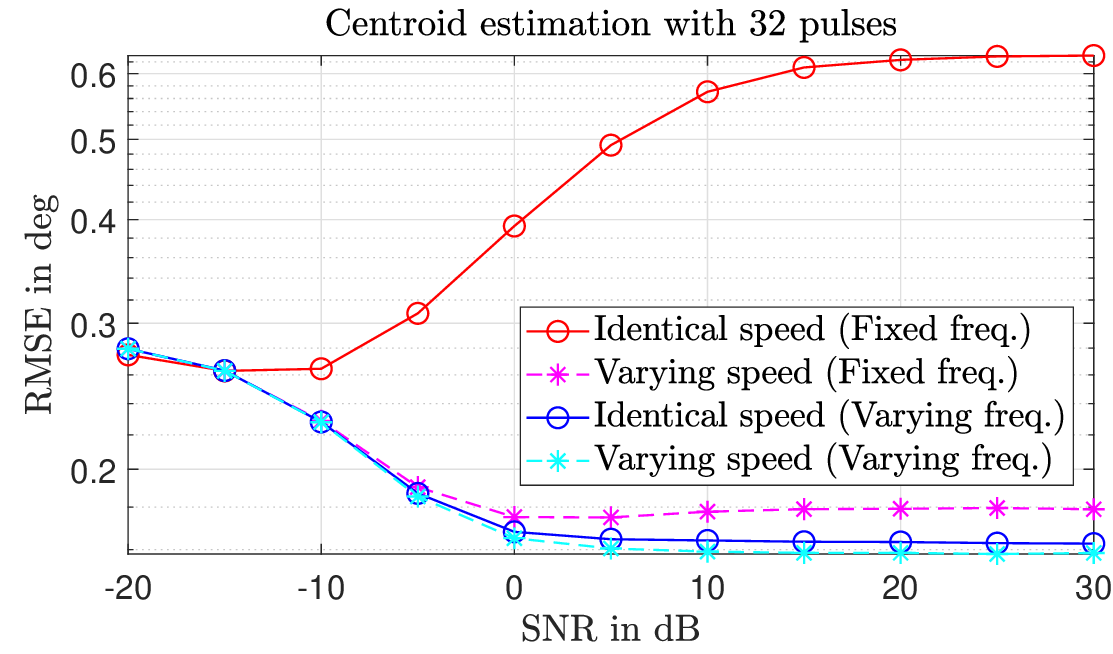}%
\label{fig:rmse_snr}}
\,
\subfloat[Comparison by the number of pulses per burst]{\includegraphics[width=.65\columnwidth]{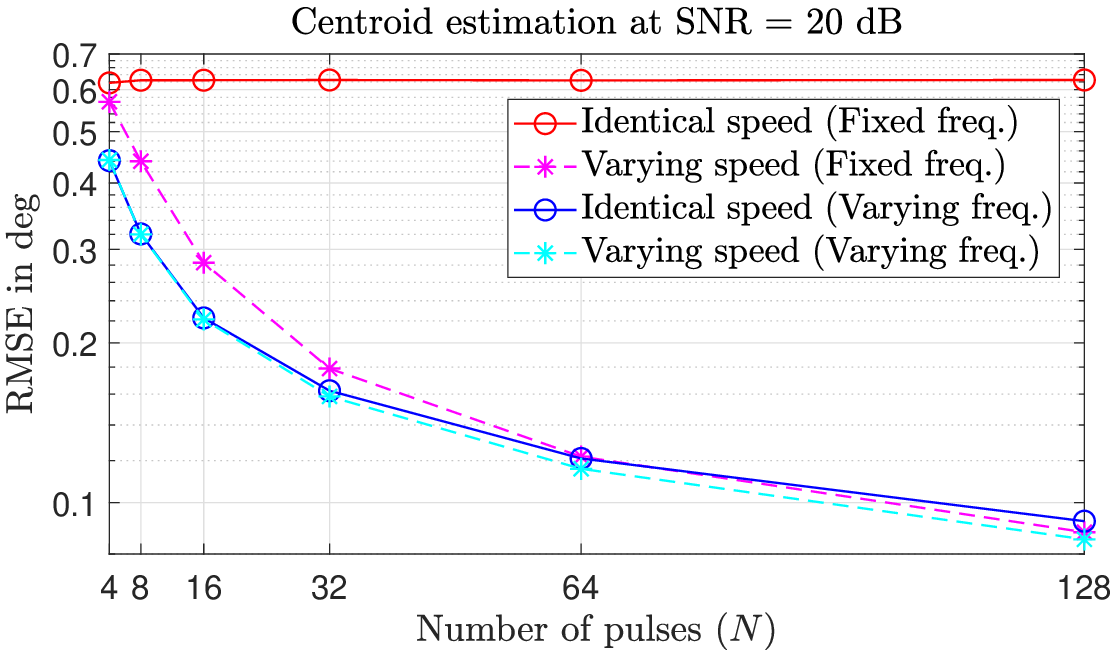}%
\label{fig:rmse_n}}
\,
\subfloat[Comparison by the number of scatterers]{\includegraphics[width=.65\columnwidth]{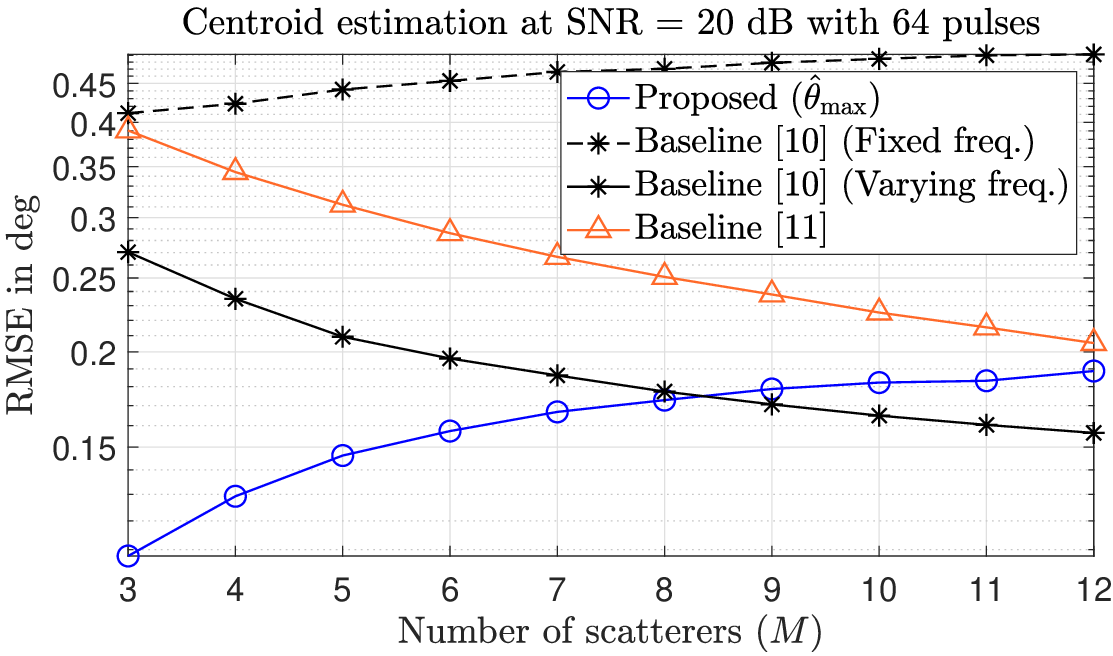}%
\label{fig:rmse_compare}}
\caption{RMSE of the centroid estimate for multiple scatterers.}
\label{fig:rmse}
\end{figure*}

Next, we present Monte Carlo simulation results to demonstrate the improved monopulse estimation performance achieved with frequency diversity.
Fig. \ref{fig:rmse} compares the root mean squared error (RMSE) of centroid estimations in various configurations.
Table \ref{tab:params} summarizes the parameters used for the simulations.
Fig. \ref{fig:rmse_snr} shows that the monopulse estimation with varying frequency can reduce RMSE compared to fixed frequency for both scatterers with identical and varying speeds.
When the pulse frequency is fixed, the RMSE tends to increase with SNR because the phases of scatterers become highly deterministic without frequency diversity, leading to biased estimates.
Fig. \ref{fig:rmse_n} shows the RMSE results by the number of pulses per burst.
The result confirms that increasing the number of pulses with different frequencies leads to accurate estimates, even when the bandwidth is fixed.
Fig. \ref{fig:rmse_compare} compares the RMSE achieved by the proposed method with the baseline approaches presented in \cite{1996trump,2012monakov}, as the number of scatterers increases under the condition that all scatterers have identical speeds.
Although the original approach in \cite{1996trump} does not employ frequency diversity, we applied it in our evaluation to ensure a fair comparison with the proposed method.
Since the method in \cite{2012monakov} relies on the amplitude-comparison monopulse algorithm, its performance remains unaffected by frequency diversity.
Despite significantly reduced computational complexity, the proposed method achieves lower RMSE than the baselines when the number of scatterers is small.
Although \cite{1996trump} and \cite{2012monakov} perform better as the number of scatterers increases, the proposed method offers a favorable trade-off between accuracy and complexity.

\begin{figure}[!h]
\centerline{\includegraphics[width=.7\columnwidth]{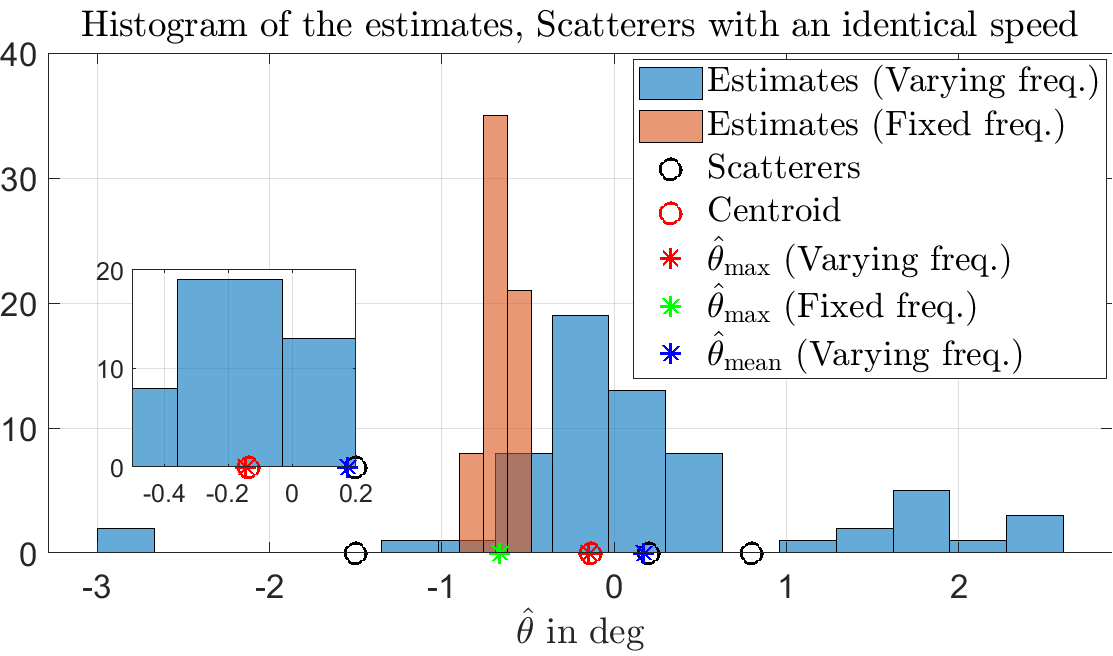}}
\caption{Histograms of estimates on 3 scatterers with an identical speed.}
\label{fig:histogram}
\end{figure}

Next, Fig. \ref{fig:histogram} presents histograms of angle estimates for 3 scatterers moving at an identical speed of 1,100 m/s where the relative radial distances are set as $\mathbf{r}=[10, -5, -7]^T$ m.
The weights and angles are configured the same as the result shown in Fig. \ref{fig:dist_Nt3}, and the SNR is set to 20 dB.
Two estimates with and without frequency diversity are compared.
To achieve frequency diversity, 64 stepped frequencies within a 150 MHz bandwidth are utilized.
The number of bins in the histogram is determined using Freedman-Diaconis rule for a large sample size, but Sturge's rule is used for a small sample size.
The mode is determined from a histogram using the following rule.
\begin{equation}
    \hat{\theta}_{\max} = \frac{h_k-h_{k-1}}{2h_k-h_{k-1}-h_{k+1}}\Delta \theta+\theta_0 +(k-1)\Delta \theta,
\end{equation}
where $k$ is the index of the peak bin, $h_k$ is the quantity of the $k$-th bin and $\Delta \theta$ is the bin width.
As depicted as the red histogram, the estimates with fixed frequency exhibit significant bias towards the false centroid due to the deterministic phases of scatterers.
However, the centroid of the scatterers can be effectively estimated as the mode of estimates when exploiting frequency diversity.
The centroid estimate becomes reliable with the adoption of frequency diversity, since the correlation between scatterer phases can be statistically relaxed.

\section{Conclusion}\label{sec:conclusion}
We proposed a centroid estimation method for multiple scatterers based on the mode of monopulse angle measurements.
The proposed method incorporates frequency diversity to mitigate correlations across measurements, thereby enhancing estimation accuracy.
We derived a semi-analytic expression of the distribution of angle estimates for multiple coherent scatterers.
Notably, we demonstrate that the mode of this distribution corresponds to the angular centroid of multiple scatterers, represented as a weighted mean of scatterer angles.
Numerical simulations validate the effectiveness of the proposed method with frequency diversity, showing that the mode-based estimator outperforms conventional techniques.

\clearpage

\bibliographystyle{IEEEtran}
\bibliography{IEEEabrv, references}

\end{document}